\documentclass[a4paper,11pt]{article}
\usepackage[utf8]{inputenc}
\usepackage{amsmath,amssymb,amsfonts}
\usepackage{cite}
\usepackage{graphicx}
\begin{document}

{\setlength{\oddsidemargin}{1.2in}
\setlength{\evensidemargin}{1.2in}}   \baselineskip 0.55cm
\begin{center}
{\bf  {P-v criticality of charged Reissner-Nordstr{\"o}m-de Sitter  black hole under the influence of Lorentz violation theory}}
\end{center}

\begin{center}
${\rm Y.\, Onika \,\, Laxmi^a}$, $\rm Ningthoujam \,\, Media^a$, $\rm T. \, Ibungochouba \,\, Singh^{a*}$ \\
$\rm {^a}Mathematics$ Department, Manipur University, Canchipur-795003 \\
*E-mail: ibungochouba@rediffmail.com
\end{center}

\begin{center}
{\bf Abstract}
\end{center}

In this paper, the modified Hawking radiation and the P-v criticality of Reissner-Nordstr{\"o}m-de Sitter (RNdS) black hole are investigated by using Dirac equation with Lorentz violation theory in curved space time. Taking cosmological constant as the thermodynamic pressure and its conjugate quantity as thermodynamic volume, the analogy between the RNdS black hole in the ensemble with fixed charge and van der Waals liquid-gas system is derived. We also study the equation of state, critical exponents, critical point, heat capacity, entropy correction, Gibbs free energy. It is worth mentioning that the critical temperature, entropy, Gibbs free energy and Helmholtz free energy are modified but the critical exponents  are found to be unchanged due to the influence of Lorentz violation theory. We observe that the number of Hawking-Page critical point increases due to Lorentz violation theory.\\



\section{Introduction}
 In 1970,  a theory based on black hole thermodynamics was proposed in order to provide a better understanding of black holes within the context of general relativity. Although this theory is still not fully understood, the black hole thermodynamics strongly reveals the fundamental and significant relationship between general relativity, thermodynamics and quantum field theory. Since it appears to be made based on the fundamental principles of thermodynamics and applies the classical and semiclassical technique to explain how the black holes describe within the domain of thermodynamics laws, its significance extends far beyond the connection between thermodynamics and black holes. The black hole thermodynamics is a significant area in gravitational physics. Hawking [1] and Bekenstein [2] developed the fundamental relationship between black holes and thermodynamics. They derived some similarities between the classical and laws of black hole thermodynamics such as the relation between black hole entropy and the area of the event horizon, the relation between mass and energy and the relation between temperature and surface gravity. These inspired Bekenstein to study the similarity between black hole entropy and the area of the event horizon. By using the quantum field theory in curved space time, the black holes behave as thermodynamic objects that transfer radiation across the event horizon, which is known as Hawking radiation. In Hawking radiation, the temperature corresponds to the surface gravity near the event horizon of black hole and entropy is equal to one-fourth of its horizon area.\\
A spherically symmetric solution for matter which represents the geometry of outside the stars and depends on the Einstein field equations was proposed by Schwarzschild [3]. By adding charge parameter, a modified form of Schwarzschild black hole was proposed by Reissner [4] in 1916. Weyl [5] and Nordstr{\"o}m [6] finalized this work which is usually known as Reissner-Nordstr{\"o}m black hole. Hawking [7] studied a theoretical explanation of black hole radiation.
As the black hole emits radiation near the event horizon, it implies that it will reduce the size of black hole and therefore increases the temperature of black hole under consideration. Hawking and Page [8] investigated the first order phase transition from thermal AdS black hole to the large AdS black hole at a certain critical temperature. Chamblin et al. [9, 10] recently studied the phase transition of first order between the small and the large  RNAdS black holes. The thermodynamic characteristics and phase transition of black holes in anti-de Sitter space (AdS) have been drawn a lot of attention particularly because of the discovery of conformal field/anti-de Sitter theory correspondence [11, 12]. By means of the AdS/conformal field theory correspondence, it was possible to explain the Hawking-Page transition as confinement/deconfinement transition in quantum chromodynamics [13].
 Over the last few decades, the  study of black hole thermodynamics has been done significantly. Many scientists have studied the thermodynamic properties of black hole within the context of thermal and non-thermal radiations [14, 15], thermodynamic phase transition [16-18], holography and geometry [19, 20] and entropy [21, 22]. These results showed that the study of black hole thermodynamics is an exciting and important subject in general relativity. Furthermore, the black holes are well-known to be thermodynamic systems with Hawking temperatures. The event horizon carries an irreducible mass distribution which is similar to thermodynamic entropy in the black hole thermodynamic system [23, 24]. Faizal and Khalil [25] investigated the thermodynamics and thermal fluctuation of  Reissner-Nordstr{\"o}m, Kerr and charged AdS black holes. Pourhassan and Faizal [26, 27] studied the thermodynamics of spinning Kerr-AdS black hole as well as its phase transition. They derived the corrected black hole entropy and found that these corrections are very effective for the geometry of small black holes.\\
Zhang [28] and Pradhan [29] studied the effect of corrected black hole entropy on the charged accelerating black hole, Kerr-Newman-AdS and RNAdS black holes. They concluded the existence of second order phase transition under the thermal fluctuations. \\
The corrected entropy and effects of logarithmic correction on the geometry of charged AdS black holes were studied by Kubiznak and Mann [30]. They showed that there is a liquid-gas type of phase transition and the critical behavior of RNdS black hole in fixed charge ensemble is also found to be consistent with that of van der Waals liquid-gas system. The thermodynamic properties for different type of black holes or black rings under quantum gravity effects have been discussed in [31-34]. Their results transformed into the original Hawking temperature to ensure reliability.
Dolan [35, 36] found a connection between the van der Waals liquid-gas system and the RNAdS black hole in extended phase space by treating the cosmological constant as thermodynamic pressure and the black hole mass as enthalpy.\\

In the past few decades, many authors studied the Lorentz violation theory based on different gravity models [37-41]. Lorentz violation theory is one of the theories which can be modified at the high energy level and  it leads to the correction of the dynamical equations for boson and fermion particles. Refs. [42-44] investigated the Lorentz violation theory in flat Euclidean space by using ether-like vectors $u^\alpha$ and Dirac equation. Liu, et al. [45] studied the entropy correction of black hole in Lorentz violation theory by applying the modified Hamilton-Jacobi equation and found that the corrected entropy depends on the choices of ether-like vectors $u^\alpha$. Refs. [46-52] discussed the modified Hawking temperature and corrected entropy of different black holes under the influence of Lorentz violation theory.\\

 The organization of this paper is as follows:
 In section 2, a brief review of Reissner-Nordstr\"{o}m-de Sitter black hole and its properties are given. The modified form of Dirac equation under Lorentz violation theory is discussed in Section 3. In Section 4, the modified heat capacity and the Hawking temperature of Reissner-Nordstr\"{o}m-de Sitter black hole are derived. The thermodynamics properties of Reissner-Nordstr\"{o}m -de Sitter black hole by taking cosmological constant as pressure are discussed in Section 5. Section 6 gives the conclusion of this paper.

\section{Reissner-Nordstr\"{o}m-de Sitter black hole }
The line element of Reissner-Nordstr\"{o}m-de Sitter (RNdS) black hole which describes the spherically symmetric charged black hole solution to Einstein-Maxwell action with positive cosmological constant  in Boyer-Lindquist coordinates, can be written as [53]
\begin{eqnarray}
ds^2=-f(r)dt^2+\frac{1}{f(r)}dr^2+r^2 d\theta^2+r^2 \sin^2\theta d\phi^2,
\end{eqnarray}
where $f(r)=1-\frac{2M}{r}+\frac{Q^2}{r^2}-\frac{\Lambda r^2}{3}$, $M$ and $Q$ represent the mass and charge of the black hole respectively and $\Lambda$ denotes the cosmological constant.
 For this RNdS black hole, the four-dimensional electromagnetic potential vector reads $A_{\mu}=(A_{\nu},0,0,0)$, and 
\begin{eqnarray}
A_{\nu}=\frac{Q}{r}.
\end{eqnarray}
The equation of event horizon for the RNdS black hole can be derived from the null hypersurface equation
\begin{eqnarray}
g^{ab}\frac{\partial F}{\partial x^{a}}\frac{\partial F}{\partial x^{b}}=0,
\end{eqnarray}
where $F$ is a function of single independent variable $r$ only. The horizon equation of RNdS black hole is given by 
\begin{eqnarray}
f(r)=1-\frac{2M}{r}+\frac{Q^2}{r^2}-\frac{\Lambda r^2}{3}=0.
\end{eqnarray}
We derive the mass of the black hole in terms of the radius of the event horizon $r_h$
 as\begin{eqnarray}
M=\frac{r_h}{2}\Big(1+\frac{Q^2}{r_{h}^2}-\frac{\Lambda r_{h}^2}{3}\Big).
\end{eqnarray}
The Hawking temperature and the heat capacity of RNdS black hole are 
\begin{eqnarray}
T&=&\frac{1}{4 \pi}\Big(\frac{\partial f}{\partial r}\Big)|_{r=r_{h}}=\frac{1}{4 \pi r_{h}}\Big(1-\frac{Q^2}{r_{h}^2}-r_{h}^2 \Lambda\Big)
\end{eqnarray}
and
\begin{eqnarray}
C&=& \Big(\frac{\partial M}{\partial T}\Big)=\frac{2 \pi r_{h}^2 \Big(Q^2-r_{h}^2+r_{h}^4 \Lambda \Big)}{r_{h}^2-3 Q^2+ r_{h}^4 \Lambda}
\end{eqnarray}respectively.

The Bekenstein-Hawking entropy of the RNdS black hole is
\begin{eqnarray}
S_{BH}=\frac{A}{4}= \pi r_{h}^2.
\end{eqnarray}

\section{ Modified Dirac equation}
The modified Dirac equation in flat space time could be obtained by applying Hamilton principle [43] under Lorentz violation theory. To obtain modified Dirac equation from flat space time to the RNdS black hole, the gamma matrices $\gamma^{\mu}$  must be derived from RNdS black hole and  the ordinary derivative should be extended to the covariant derivative. The dynamical equation of fermions with spin-$1/2$ particles under the Lorentz violation theory is given by [54]
\begin{eqnarray}
\gamma^{\mu}D_{\mu}[1+\hbar^2\frac{a}{m^2}(\gamma^{\mu}D_{\mu})^2]\Psi+\big[\frac{b}{\hbar}\gamma^{5}+c\hbar(u^{\mu}D_{\mu })^2-\frac{m}{\hbar}\big]\Psi=0,
\end{eqnarray}
where $a,b$ and $c$  are small quantities. $\Psi$ and $\hbar$ are the wave function and Planck constant respectively. $u^{\mu}$ represents  the ether-like vectors that satisfy the condition $u^{\mu}u_{\mu}={\rm constant}.$
The operators $D_{\mu}$ and $\Pi_{\alpha \beta}$ are defined by
\begin{eqnarray}
D_{\mu}&=& \partial_{\mu}+\frac{i}{\hbar}q A_{\mu}+\frac{i}{2}\Gamma^{\alpha \beta}_{\mu}\Pi_{\alpha \beta},\cr
\Pi_{\alpha \beta}&=&\frac{i}{4}[\gamma^{\alpha},\gamma^{\beta}].
\end{eqnarray}
Here,  $\Pi_{\alpha \beta}$ represents the Lorentz spinor generator, $\Gamma^{\alpha \beta}_{\mu}$ are the spin connections and $\frac{i}{2}\Gamma^{\alpha \beta}_{\mu}\Pi_{\alpha \beta}$ are the rotational contact terms. It is noted that the contact terms could be ignored in semiclassical theory.
The gamma matrices for the RNdS black hole can be constructed as follows
\begin{eqnarray}
 \gamma^{t}&=& \frac{1}{\sqrt{-f(r)}}
 \begin{bmatrix}
i&0\\
0&-i\\
\end{bmatrix},\nonumber\\
 \gamma^{r}&=&\sqrt{\frac{1}{f(r)}}
 \begin{bmatrix}
0&\sigma^3\\
\sigma^3&0\\
\end{bmatrix},\nonumber\\
  \gamma^{\theta}&=&\frac{1}{r}
  \begin{bmatrix}
0&\sigma^1\\
\sigma^1&0\\
\end{bmatrix},\nonumber\\
   \gamma^{\phi}&=&\frac{1}{r \sin\theta}
   \begin{bmatrix}
0&\sigma^2\\
\sigma^2&0\\
\end{bmatrix},
\end{eqnarray}
where $\sigma^{1},\sigma^{2}$ and $\sigma^{3}$ are Pauli matrices satisfying the condition
\begin{eqnarray}
\gamma^{\alpha} \gamma^{\beta}+\gamma^{\beta} \gamma^{\alpha}&=&\lbrace\gamma^{\alpha},\gamma^{\beta}\rbrace =2g^{\alpha \beta} I.
\end{eqnarray}
From Eq. (11), the relation between $\gamma^5$ and $\gamma^{\mu}$ can be written as 
\begin{eqnarray}
\gamma^5 \gamma^{\mu}+\gamma^{\mu}\gamma^5 &=&0.
\end{eqnarray}
The expression of $\gamma^5$ for RNdS black hole is derived as 
\begin{eqnarray}
\gamma^5 &=& i \gamma^t \gamma^r \gamma^{\theta}\gamma^{\phi}=\Big(f(r)r^2 \sin \theta\Big)^{-1}
 \begin{bmatrix}
0&-1\\
1&0\\
\end{bmatrix}.
\end{eqnarray}
The wave function $\psi$ for fermion with spin-$1/2$ particles in accordance with semiclassical theory can be written as
\begin{eqnarray}
\psi=\begin{pmatrix} \alpha \\ \beta \end{pmatrix} {\rm exp}\left[\frac{i}{\hbar} S \right],
\end{eqnarray}
where $S$ denotes the action of Dirac particles. Using Eq. (15) in Eq. (9), the modified equation is derived as
\begin{eqnarray}
g^{\mu \nu}(\partial_{\mu}S+q A_{\mu})(\partial_{\nu}S+q A_{\nu}) (1+2 a_{k})+\lambda u^\mu u^\nu (\partial_{\mu}S+q A_{\mu})(\partial_{\nu}S+q A_{\nu})+m^2=0.
\end{eqnarray}
The above equation indicates the dynamical equation for a Dirac particles with spin-$1/2$, mass $m$ and charge $q$.
Our aim is to solve the action $S$ from Eq. (16) and the corresponding corrected entropy, Hawking temperature and thermodynamic properties can be studied in accordance with the semiclassical theory and the WKB approximation.
\section{Derivation of modified Hawking temperature for RNdS black hole}
For solving Eq. (16), we must construct ether-like vectors $u^{\alpha}$ from Eq. (1). It is noted that for the flat space time, all the components of $u^{\alpha}$ are constant vectors and automatically hold the condition $u^{\alpha}u_{\alpha}={\rm constant}$. In the case of curved spacetime, all the components of $u^{\alpha}$ are not constant vectors but we can construct 
\begin{eqnarray}
u^{\alpha}u_{\alpha}= {\rm constant}.
\end{eqnarray}
From Eq. (1), the ether-like vectors for the RNdS black hole are constructed as follows
\begin{eqnarray}
u^t &=&\frac{c_{t}}{\sqrt{f(r)}}=\frac{c_{t}}{\sqrt{1-\frac{2M}{r}+\frac{Q^2}{r^2}-\frac{\Lambda r^2}{3}}},\cr
u^r &=&c_{r} \sqrt{f(r)}=c_{r}{\sqrt{1-\frac{2M}{r}+\frac{Q^2}{r^2}-\frac{\Lambda r^2}{3}}},\cr
u^\theta &=&\frac{c_{\theta}}{\sqrt{g_{\theta\theta}}}=\frac{c_{\theta}}{r},\cr
u^\phi &=&\frac{c_{\phi}}{\sqrt{g_{\phi\phi}}}=\frac{c_{\phi}}{r \sin \theta},
\end{eqnarray}
where $c_t, c_r, c_{\theta}$ and $c_{\phi}$ are real constants. Eq. (18) also satisfies  $u^\alpha u_{\alpha}= -c_t^2+c_{r}^2+c_{\theta}^2+c_{\phi}^2 =$ constant. 
Using Eqs. (1) and (18) in Eq. (16), we get

\begin{eqnarray}
&&\Big[g^{tt}(\frac{\partial S}{\partial t}+q A_{t})^2+g^{rr}(\frac{\partial S}{\partial r})^2+g^{\theta \theta}(\frac{\partial S}{\partial \theta})^2+g^{\phi \phi}(\frac{\partial S}{\partial \phi})^2\Big](1+2 a_{k})\cr&&+\lambda u^t u^t (\frac{\partial S}{\partial t}+q A_{t})^2+ \lambda u^r u^r (\frac{\partial S}{\partial r})^2+\lambda u^\theta u^\theta (\frac{\partial S}{\partial \theta})^2+\lambda u^\phi u^\phi (\frac{\partial S}{\partial \phi})^2\cr&& +2\lambda u^t u^r (\frac{\partial S}{\partial t}+q A_{t})(\frac{\partial S}{\partial r})+2\lambda u^t u^\theta (\frac{\partial S}{\partial t}+q A_{t})(\frac{\partial S}{\partial \theta})+2\lambda u^t u^\phi (\frac{\partial S}{\partial t}+q A_{t})(\frac{\partial S}{\partial \phi})\cr&&+2\lambda u^r u^\theta (\frac{\partial S}{\partial r})(\frac{\partial S}{\partial \theta})+2\lambda u^r u^\phi (\frac{\partial S}{\partial r})(\frac{\partial S}{\partial \phi})+2\lambda u^\theta u^\phi (\frac{\partial S}{\partial \theta})(\frac{\partial S}{\partial \phi})+m^2=0.
\end{eqnarray}
The above equation contains the variables $t, r, \theta$ and $\phi$. To investigate the Hawking radiation of black hole, we take the action $S$ as
\begin{eqnarray}
S=-\omega t+X(r)+Y(\theta)+Z(\phi)+\delta,
\end{eqnarray}
where $\omega$ denotes the energy of the particle and
$\frac{\partial S}{\partial t}=-\omega,\,\,
\frac{\partial S}{\partial r} = \frac{\partial X}{\partial r},\,\,
\frac{\partial S}{\partial \theta}=\frac{\partial Y}{\partial \theta},\,\,
\frac{\partial S}{\partial \phi}=\frac{\partial Z}{\partial \phi}.$ $\delta$ is a complex constant.
Substituting Eq. (20) in Eq. (19), we derive a quadratic equation in $
\frac{\partial X}{\partial r}$ as
\begin{eqnarray}
A\Big(\frac{\partial X}{\partial r}\Big)^2 +B\Big(\frac{\partial X}{\partial r}\Big) +C =0,
\end{eqnarray}
where $A$, $B$ and $C$ are defined by
 \begin{eqnarray}
 A&=&f(r)(1+2 a_k +\lambda c_{r}^2),\cr
 B&=&2\lambda c_{t} c_{r} (-\omega+q A_t)+2\lambda c_{r} c_{\theta} \frac{\sqrt{f(r)}}{r}\Big(\frac{\partial Y}{\partial \theta}\Big)+2\lambda c_r c_{\phi}\frac{\sqrt{f(r)}}{r \sin \theta}\Big(\frac{\partial Z}{\partial \phi}\Big),\cr
 C&=&\left[\frac{-1}{f(r)}(-\omega +q A_t)^2 +\frac{1}{r^2}\Big(\frac{\partial Y}{\partial \theta}\Big)^2+\frac{1}{r^2 \sin^2\theta}\Big(\frac{\partial Z}{\partial \phi}\Big)^2\right](1+2a_k)\cr&&+\frac{\lambda c_t^2}{f(r)}(-\omega +q A_t)^2+\frac{2\lambda c_t c_{\theta}}{r\sqrt{f(r)}}(-\omega +q A_t)\left(\frac{\partial Y}{\partial \theta}\right)\cr&&+\frac{2\lambda c_t c_{\phi}}{\sqrt{f(r)}r\sin \theta}(-\omega +q A_t)\Big(\frac{\partial Z}{\partial \phi}\Big)+\frac{\lambda c_{\theta}^2}{r^2}\Big(\frac{\partial Y}{\partial \theta}\Big)^2+\frac{2\lambda c_{\theta} c_{\phi}}{r^2 \sin\theta}\left(\frac{\partial Y}{\partial \theta}\right)\Big(\frac{\partial Z}{\partial \phi}\Big)\cr&&+\frac{\lambda c_{\phi}^2}{r^2 \sin^2\theta}\Big(\frac{\partial Z}{\partial \phi}\Big)^2+m^2.
\end{eqnarray}
Using Eq. (21) in Eq. (20), the action $S$ can be expressed as\begin{eqnarray}
S&=&-\omega t+\int\frac{-B\pm \sqrt{B^2-4A C}}{2A}+Y(\theta)+Z(\phi)+\delta.
\end{eqnarray}
Calculating the above integral across the event horizon of RNdS black hole by using Feynman prescription and residue theorem of complex analysis, we derive the imaginary part of the radial action as
\begin{eqnarray}
X(r)_{\pm}=\frac{i \pi 3r_{h}^3(\omega-qA_t)L}{2(3Mr_{h}-3Q^2-\Lambda r_{h}^4)},
\end{eqnarray}
where $L=\frac{\lambda c_t c_r \pm \sqrt{(1+2a_k)[1+2a_k +\lambda (c_{r}^2-c_{t}^2)]}}{1+2a_k +\lambda c_r^2}.$
$X_{+}(r)$ and $X_{-}(r)$ are the outgoing and ingoing solutions respectively. According to semiclassical approximation, the emission rate of fermions with spin-$1/2$ particle across the event horizon of the RNdS black hole can be written as
\begin{eqnarray}
\Gamma_{emission}&=&{\rm exp}(-2 {\rm Im} S_{+})={\rm exp}[-2({\rm Im} X_{+}(r)+{\rm Im} \delta)]
\end{eqnarray} and
\begin{eqnarray}
\Gamma_{absorption}&=&{\rm exp}(-2 {\rm Im} S_{- })={\rm exp}[-2({\rm Im} X_{-}(r)+{\rm Im} \delta)].
\end{eqnarray}
If ${\rm Im \delta}=-{\rm Im X_{-}}$, then the probability is normalized. In such case, there is $100\%$ chance of ingoing fermions with spin-$1/2$ particle entering the black hole in accordance with semiclassical WKB approximation. Since $X_{+}(r)=-X_{-}(r)$, the tunneling of fermions with spin-$1/2$  particles from inside to outside across the event horizon of RNdS black hole is given by
\begin{eqnarray}
 \Gamma_{rate}&=&\frac{\Gamma_{emission}}{\Gamma_{absorption}}\cr
&=& {\rm exp}\left[\frac{-6\pi r_{h}^3(\omega-q A_t)L_1}{3Mr_{h}-3Q^2-\Lambda r_{h}^4}\right],
\end{eqnarray}
which is similar to the Boltzmann factor; ${\rm exp}(-\omega \beta)$, where $\beta$ represents the inverse temperature of the black hole. The Hawking temperature near the event horizon of RNdS black hole is derived as
\begin{eqnarray}
T_H &=&\frac{1}{4 \pi r_{h} L_1}\Big(1-\frac{Q^2}{r_{h}^2}-r_{h}^2 \Lambda\Big) \cr
&=&\frac{T}{L_1},
\end{eqnarray}
where  $L_1=\frac{ \sqrt{(1+2 a_k)[1+2a_k +\lambda(c_{r}^2-c_{t}^2)]}}{1+2a_k +\lambda c_r^2}$.
\begin{figure}[h]
\centering
\includegraphics{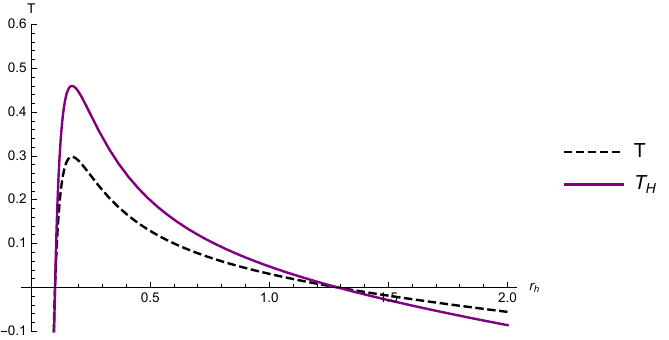}
\caption{Plot of original and modified Hawking temperatures of RNdS black hole with radius of event horizon, ${r_h}$. Here, $a_k=0.6,\Lambda=0.6, c_t=1, c_r=0.2, \lambda=0.7$ and $Q=0.1.$}
 \end{figure}
It is noted from Eqs. (6) and (28) that the Hawking temperature of RNdS black hole is modified under Lorentz violation theory. If $0<L_{1}<1$, then $T_H>T$. When $1<L_1<\infty$, $T_H<T$. If $L_1=1$, the modified Hawking temperature, $T_H$ tends to the original Hawking temperature, $T$. The variation of $T_H$ and $T$ with the radius of event horizon is shown in Fig. 1 with the set of black hole parameters. It is noted from the Fig. 1 that the modified Hawking temperature is larger than the original Hawking temperature. The maximum value of modified temperature $T_H$ and original temperature $T$ are $0.459828$ and $0.29806$ respectively when $r_h=0.17169.$
The two Hawking temperatures, $T$ and $T_H$ intersect at the points $r_h=0.100303$ and $r_h=1.28709$ respectively. If $0.100303<r_h<1.28709$, then the inequality holds $T<T_H$. When $1.28709<r_h<\infty$, then it follows the inequality $T>T_H$.
 
 The heat capacity of RNdS black hole under Lorentz violation is given by 
\begin{eqnarray}
 C_H&=&\frac{\partial M}{\partial T_H}\cr
 &=&\frac{2 \pi L_1 r_{h}^2 (Q^2-r_{h}^2+r_{h}^4 \Lambda)}{r_{h}^2-3 Q^2 +r_{h}^4 \Lambda}.
\end{eqnarray}
From Eqs. (7) and (29), the heat capacity of RNdS black hole is modified due to Lorentz violation theory. If $L_1$ tends to unity, the modified heat capacity $C_H$ approaches to the original heat capacity $C$. If $0<L_1<1$, the heat capacities obey the inequality $C_H<C$. When $1<L_1<\infty$, they hold the inequality $C_H>C$. The change of modified heat capacity from original heat capacity with the radius of event horizon is shown in Fig. 2. The RNdS black hole has a phase transition when $r_h=0.171693$. The RNdS black hole is stable if $0<r_h<0.171693$ and is unstable if $0.171693<r_h<\infty$ for the above set of parameters. We observe that the position of phase transition is not affected by the influence of Lorentz violation theory.

 \begin{figure}[h]
\centering
\includegraphics{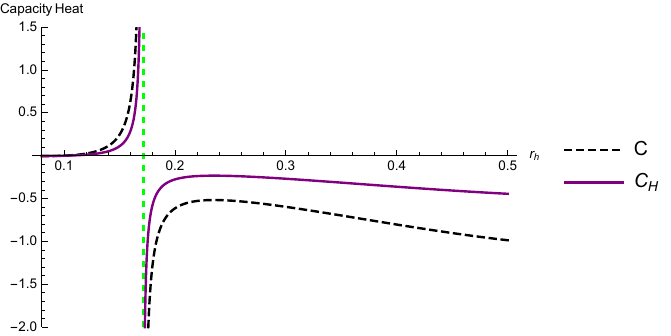}
\caption{Plot of original and modified heat capacity of RNdS black hole with radius of event horizon, ${r_h}$. Here, $\Lambda=1, a_k=0.1, c_t=1, c_r=2, \lambda=0.9$ and $Q=0.1.$}
 \end{figure}
 \section{Thermodynamics of RNdS black hole}
 For the assymtotically de Sitter black hole in four dimensions, the thermodynamic pressure is related to the cosmological constant for the RNdS black hole [38] as
\begin{eqnarray}
P&=&-\frac{\Lambda}{8 \pi}.
\end{eqnarray}
Then the modified Hawking temperature $T_H$ can be expressed in terms of thermodynamic pressure as follows
\begin{eqnarray}
T_{H}&=&\frac{1}{4 \pi r_h L_1}\Big(1+8\pi P r_{h}^2-\frac{Q^2}{r_{h}^2}\Big).
\end{eqnarray}

\begin{figure}[h!]
\centering
\includegraphics{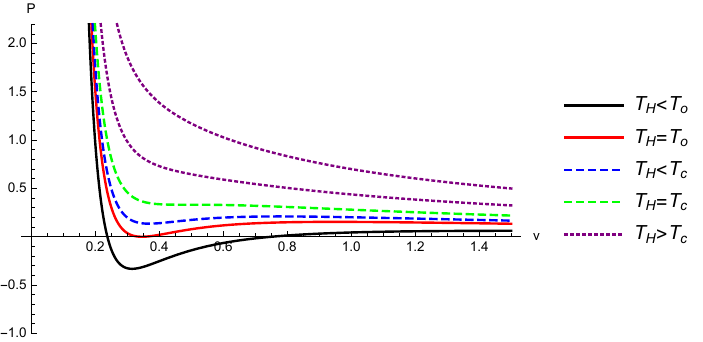}
\caption{ P-v diagram}
\end{figure}

Eq. (31) can be transformed into the equation of state in the extended phase space for a fixed charge of RNdS black hole, $P=P(V,T_H)$,
\begin{eqnarray}
P=
\frac{Q^2}{8\pi r_{h}^4}-\frac{1}{8 \pi r_{h}^2}+\frac{L_1 T_{H}}{2r_{h}},\,\,\,\,  r_{h}=\left(\frac{3V}{4\pi}\right)^{\frac{1}{3}},
\end{eqnarray}
where $T_H$ and $Q$ indicate the modified Hawking temperature and its charge of RNdS black hole respectively. $V$ represents the thermodynamic volume, which is given in terms of event horizon $r_h$.
Replacing $2 r_{h}$ by $v$ in Eq. (32), we get
\begin{eqnarray}
P=\frac{2Q^2}{\pi v^4}-\frac{1}{2\pi v^2}+\frac{L_1 T_H}{v}.
\end{eqnarray}
Since $L_1=L_1(a_k,\lambda, c_t, c_r)$ and the presence of Lorentz violation parameter $\lambda$ in the above equation, the expression of thermodynamic pressure is affected. We see from Eqs. (31) and (33) that the Hawking temperature and the thermodynamic pressure are modified due to Lorentz violation theory. If $L_1>1$ or $L_1<1$, the modified thermodynamic pressure is greater or less than original thermodynamics pressure derived in [30].  If $L_1=1$ in Eq. (31) and Eq. (33), we get the original Hawking temperature given in Eq. (6) and thermodynamic pressure given in [30]. The $P-v$ diagram is drawn in Fig. 3 and if $Q\neq 0$ and $T_H<T_c$, there exists an inflexion point and the behavior is similar to the van der Waals equation of liquid-gas system. To investigate the thermodynamic behavior for RNdS black hole, the critical volume and the pressure are calculated through the equations
\begin{eqnarray}
\frac{\partial P}{\partial v}\Big {|}_{T_H=T_c}=0,\,\,\
\frac{\partial^2 P}{\partial v^2}\Big {|}_{T_H=T_c}=0.
\end{eqnarray}

\begin{figure}[h!]
\centering
\includegraphics[width=205pt,height=170pt]{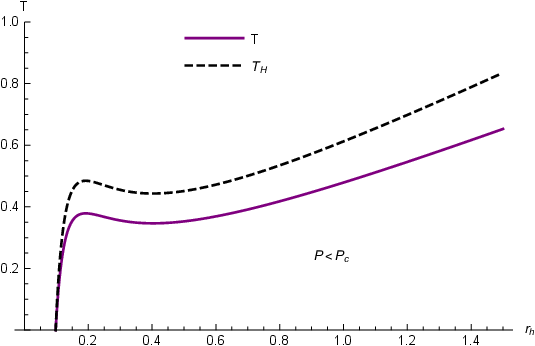}
\hfill
\includegraphics[width=205pt,height=170pt]{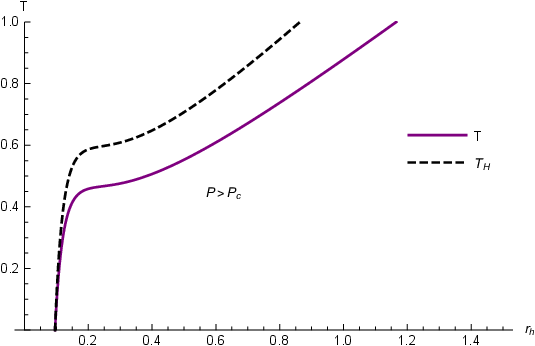}
\caption{\label{fig 3} Modified and original Hawking temperatures with the radius of event horizon $r_h$ when $P>P_c$ (right panel) and $P<P_c$ (left panel).}
\end{figure}
From Eqs. (33) and (34), the critical points are derived as
\begin{eqnarray}
P_c=\frac{1}{96 \pi Q^2},\,\,
v_c=2\sqrt{6}Q,\,\,
T_c = \frac{1}{3\sqrt{6}L_1 \pi Q}.
\end{eqnarray}
From Eq. (35), we see that the critical pressure as well as critical volume is not affected by the influence of Lorentz violation theory but the critical temperature is affected. The critical temperature of RNdS black hole may increase or decrease according to $0<L_1<1$ or $1<L_1<\infty $. If $L_1=1$, we get original critical temperature. From Eq. (34) and $P=0$, we get a temperature $T_0$ which is similar to the van der Waal equation as
\begin{eqnarray}
T_0=\frac{1}{6\sqrt{3}L_{1}\pi Q}.
\end{eqnarray}
 We see that the temperature $T_0$ of RNdS black hole depends on the charge, $Q$ and Lorentz violation parameter, $\lambda$ but the temperature $T_0$ is inversely proportional to charge $Q$ in the absence of Lorentz violation theory. We also see from the Fig. 3 that the pressure is negative for some $r_h$ if $T_H<T_0$. This behavior is unphysical related to fluid. Therefore, it needs to be replaced the oscillatory part of isotherm by an isobar in accordance with Maxwell's area law [56].  The modified Hawking temperature which is drawn in Fig. 4 at $P<P_c$ or $P>P_c$ is always greater than the original Hawking temperature due to the influence of Lorentz violation theory.
The universal constant of RNdS black hole in Lorentz violation theory can be calculated by using these critical quantities given in Eq. (35) as
 \begin{eqnarray}
\rho= \frac{P_c v_c}{T_c}&=& \frac{3L_1}{8}.
\end{eqnarray}
It is noted that the universal constant of RNdS black hole in Lorentz violation theory comparing with a charged de Sitter black hole in Einstein gravity appears a constant term $L_1$ related to Lorentz violation. This constant term $L_1$ is unity in Einstein gravity. The critical exponents of RNdS black hole in Lorentz violation theory can also be derived. Let us define
\begin{eqnarray}
 p=\frac{P}{P_c},\,\, \nu=\frac{v}{v_c},\,\,\tau=\frac{T_H}{T_c},\,\,t=\tau -1.
 \end{eqnarray}
Using Eq. (38) in Eq. (33), the equation of state can be written as
\begin{eqnarray}
8\tau&=& 3\nu (p+\frac{2}{\nu^2})-\frac{1}{\nu^3},
\end{eqnarray}
which is consistent with a charged AdS black hole in the Einstein gravity [30].
The expressions of the four critical exponents $\bar{\alpha}, \bar{\beta},\bar{\gamma}$ and $\bar{\delta}$ are defined as follows:\\
$\bar{\alpha}$ governs the behavior of specific heat at constant volume,
\begin{eqnarray}
C_v&=&T_H\frac{\partial S_{bhL}}{\partial T_H}\Big{|}_v\propto\big|t\big|^{-\bar{\alpha}}.
\end{eqnarray}
$\bar{\beta}$ is related to the behaviour of the order parameter $\eta$ which is given by
\begin{eqnarray}
\eta&=&v_g-v_l\propto \big|t\big|^{\bar{\beta}}.
\end{eqnarray}\
$\bar{\gamma}$ is related to the behavior of isothermal compressibility $k_T$ as
\begin{eqnarray}
k_T\propto \big|t\big|^{\bar{\gamma}}.
\end{eqnarray}
$\bar{\delta}$ describes the variation of pressure relative to volume as
\begin{eqnarray}
P-P_c\propto \big|v-v_c\big|^{\bar{\delta}}. 
\end{eqnarray}
After some calculations, the critical exponents are derived as $\bar{\alpha}=0,\,\,\bar{\beta}=\frac{1}{2},\,\,\bar{\gamma}=1,\,\,\bar{\delta}=3$.
This shows that the critical exponents of RNdS black hole in Lorentz violation theory are the same as for a charged AdS black hole in the Einstein gravity [55]. Hence, the Lorentz violation theory does not change the values of critical exponents. We conclude that the equation of state of RNdS black hole is dependent on the Lorentz violation parameter  $\lambda$ but the law of corresponding state of RNdS black hole is independent of Lorentz violation parameter $\lambda$ which is the same in Einstein gravity where $\lambda=0$. This results imply that the critical exponents are equal to the Einstein gravity.\\
We also see that the Hawking temperatures $T$ and $T_H$ have the same critical pressure at $P_c=\frac{1}{96 \pi Q^2}$. The graphs of $T$ and $T_H$ with the event horizon of RNdS black hole are drawn in Fig. 4. If $P>P_c$, the temperatures $T$ and $T_H$ are monotonic functions with respect to black hole radius, $r_h$. When $P<P_c$, then $\frac{\partial T_H}{\partial r_h}=0$ gives the local maximum and local minimum values of the modified Hawking temperature. Then we derive the local minimum and local maximum of Hawking temperature as follows
\begin{eqnarray}
T_{H (min/max)}&=&\frac{2\sqrt{P}}{L_1\sqrt{\pi}}\frac{\Big(1-32\pi PQ^2 \pm\sqrt{1-96 \pi PQ^2}\Big)}{(1\pm \sqrt{1-96 \pi PQ^2})^\frac{3}{2}}.
\end{eqnarray}
We see from Eq. (44) that there are three branches of black hole solutions namely small, $T_{H (min)}$, intermediate, $T_H$ and large black hole, $T_{H (max)}$. There exists a first order phase transition from small black hole to the large black hole which is similar to the van der Waals equation of liquid-gas system. The black hole is unstable for the intermediate solution. It is known from the graphs of Fig. 4 that the modified Hawking temperature is greater than the original Hawking temperature due to the influence of Lorentz violation theory.
\subsection{Entropy correction}
In this section, our aim is to investigate the entropy correction for RNdS black hole under the influence of Lorentz violation theory. Refs. [57-60] studied the corrected entropy of black hole by applying null geodesic technique and the particle back reaction effects. Refs. [61, 62] also investigated the corrected Hawking temperature and entropy of black hole using first law of black hole thermodynamics. Applying beyond semiclassical method, we also calculate the Bekenstein-Hawking entropy for RNdS black hole under the Lorentz violation theory as
\begin{figure}[h]
\centering
\includegraphics{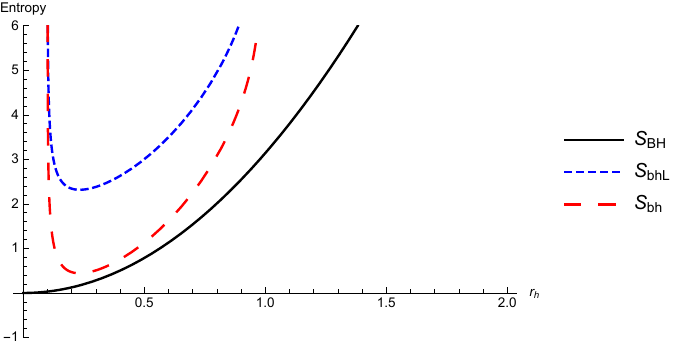}
\caption{Plot of original entropy $(S_{BH})$, corrected entropy under the influence of Lorentz violation $(S_{bhL})$ and corrected entropy beyond the semiclassical approximation $(S_{bh})$ with the radius of event horizon, $r_h$. Here $\Lambda=1, a_k=1, c_t=0.6, c_r=1.2, \lambda=0.3$ and $Q=0.1$.}  
 \end{figure}
  
\begin{eqnarray}
S_{bhL}&=&S_{BH}-log\left(1-Q^2/r_{h}^2-\Lambda r_{h}^2 \right)+\frac{1}{2}log(16 L_1^2 \pi)\cr
&=&S_{bh}+\frac{1}{2}log(16 L_1^2 \pi),\end{eqnarray}
where $S_{bh}=S_{BH}-log\left(1-Q^2/r_{h}^2-\Lambda r_{h}^2 \right)$.
$S_{bh}$ and $S_{BH}$ denote the Bekenstein-Hawking entropy beyond semiclassical approximation and original Bekenstein-Hawking entropy of RNdS black hole. We see from Eq. (45) that the effects of Lorentz violation theory causes the increase of black hole entropy. In the absence of Lorentz violation theory, the presence of logarithmic term prevents the rise of Bekenstein-Hawking entropy.

In Fig. 5, we draw the Bekenstein-Hawking entropy with the radius of event horizon, $r_h$ for the set of black hole parameters. The path of original Bekenstein-Hawking entropy $(S_{BH})$ is a parabola passing through origin whereas the paths of $S_{bhL}$ and $S_{bh}$ are semilike parabolas due to the Lorentz violation theory and quantum gravity effects beyond the semiclassical approximation respectively.

 The two paths $S_{bhL}$ and $S_{bh}$ coincide in the absence of Lorentz violation theory. If the logarithmic terms tend to zero, $S_{bhL}$ and $S_{bh}$ tend to the original Bekenstein-Hawking entropy of RNdS black hole.
 
 \subsection{Gibbs free energy}
 The Gibbs free energy can be used to investigate the maximum of reversible work done by a thermodynamic system in the extended phase space. It is noted that the black holes are stable or unstable if Gibbs free energy are positive or negative. The real roots of $G(r_h)$ indicate the location of Hawking-Page phase transition points.  If the Gibbs free energy is negative, the black hole is in radiative phase. To investigate the global stability and the Hawking-Page phase transition points, we need to calculate the Gibbs free energy of RNdS black hole. The Gibbs free energy is defined as  $G=M-T_{H} S_{bh}-A_{\mu}Q$ [63]. Then the original and the modified Gibbs free energy of RNdS black hole are 
 \begin{eqnarray}
G_{BH}=\frac{r_h}{4}\Big(1-\frac{Q^2}{r_h^2}+\frac{\Lambda r_h^2}{3}\Big)
\end{eqnarray}
and
 \begin{eqnarray}
G_{bh}
&=& \frac{r_h}{2}\left(1-\frac{Q^2}{r_h^2}-\frac{\Lambda r_h^2}{3}\right)-\frac{1}{4L_1 \pi r_h}\left(1-\frac{Q^2}{r_h^2}-\Lambda r_h^2\right)\Big[\pi r_h^2+\frac{1}{2}log (16 \pi L_1^2)\cr&&-log\Big(1-\frac{Q^2}{r_h^2}-\Lambda r_h^2\Big)\Big]\cr&&
\end{eqnarray}
respectively. It is noted from Eqs. (46) and (47) that the original Gibbs free energy $(G_{BH})$ is larger than the modified Gibbs free energy $(G_{bh})$ due to presence of logarithmic correction term and Lorentz violation theory.
The event horizon of original Gibbs free energy can take all values from zero to infinity but the range of event horizon of modified Gibbs free energy depends on charge of the black hole and cosmological constant due to the presence of logarithmic term.\\
In Figs. (6), (7) and (8), we see the behaviors of original and modified Gibbs free energy and obtain the effects of Lorentz violation theory. The variation in the graphs of modified Gibbs free energy from the original Gibbs free energy is due to the Lorentz violation theory. The original Gibbs free energy tends to converge at equilibrium point in the stable region of RNdS black hole for different values of charge and fixed cosmological constant as shown in Fig. 6(a). For different values of charge $Q$, with fixed $\lambda=0.4, a_k=0.1, c_t=1.5, c_r=1$ and $\Lambda=0.2$, the modified Gibbs free energy converges to the point $r_h=1.14$ as shown in Fig. 7(a). Hence RNdS black hole undergoes Hawking-Page phase transition point at $r_h=1.14$. Above the Hawking-Page phase transition point, the black hole is stable and below the Hawking-Page phase transition point, the black hole is unstable as shown in Fig. 7(a). The stability and the minimum value of modified Gibbs free energy are shown in Table 1. For fixed charge, $Q=0.1$ and different values of cosmological constant $\Lambda$, the original Gibbs free energy has a Hawking-Page phase transition point at $r_h\simeq 0.1$. The black hole is stable in the region $0.1<r_h<\infty$ and unstable in the region $0<r_h<0.1$ which is shown in Fig. 6(b). For different values of cosmological constant $\Lambda$ and fixed value of $Q=0.1,\lambda=0.4, a_k=0.1, c_t=1.5, c_r=1$, the modified Gibbs free energy has a Hawking-Page phase transition point at $r_h=0.1001$. Then the modified Gibbs free energy decreases monotonically for different values of cosmological constant. After attaining the minimum values, the Gibbs free energy diverges at different Hawking-Page transition points due to Lorentz violation theory as shown in Fig. 7(b). The range of stability and minimum value of Gibbs free energy are shown in Table 2. 
We also discuss the behavior of Gibbs free energy with event horizon $r_h$ for different values of Lorentz violation parameter $\lambda$ and fixed value of $Q=0.1, \Lambda=0.2, a_k=0.1,c_t=1.5, c_r=1$ in Fig. 8. The Gibbs free energy decreases from the Hawking-Page transition point $r_h=0.1001$ to the negative value monotonically. After reaching minimum values, it converges to a state of equilibrium for different values of Lorentz violation parameter $\lambda$. The proof of the  statements is shown in Table 3. From Figs. 6, 7 and 8, the influence of Lorentz violation theory increases the number of Hawking-Page transition point for RNdS black hole. The graphs of $7(a)$ and $7(b)$ are consistent with $6(a)$ and $6(b)$ respectively in the absence of logarithmic term and Lorentz violation theory. This shows that the Gibbs free energy is affected due to Lorentz violation theory.
  \begin{figure}[h!]
\centering
\includegraphics[width=205pt,height=170pt]{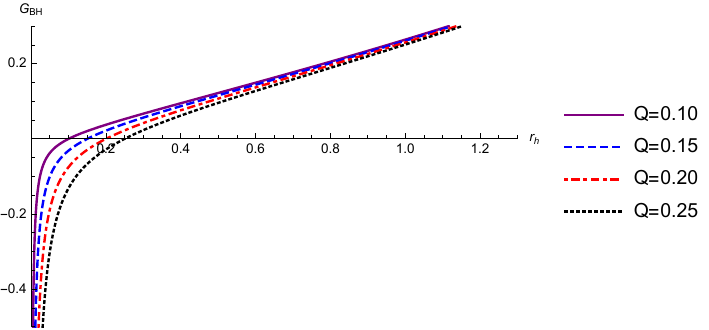}
\hfill
\includegraphics[width=205pt,height=170pt]{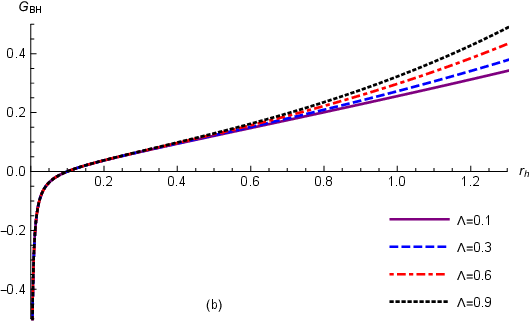}
\caption{$G_{BH}$ vs $r_h$ with varying charge $Q$ with fixed $\Lambda=0.2$ (left plot) and $G_{BH}$ vs $r_h$ with varying parameter $\Lambda$ (right plot) with fixed $Q=0.1$.}
\end{figure}
   \begin{figure}[h!]
\centering
\includegraphics[width=205pt,height=170pt]{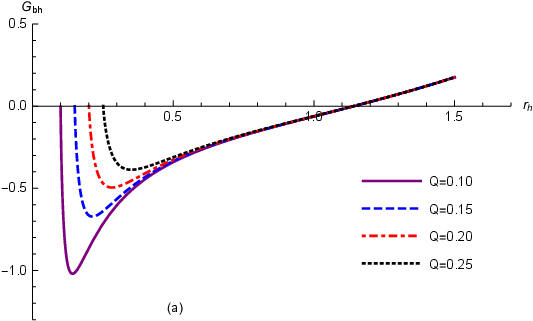}
\hfill
\includegraphics[width=205pt,height=170pt]{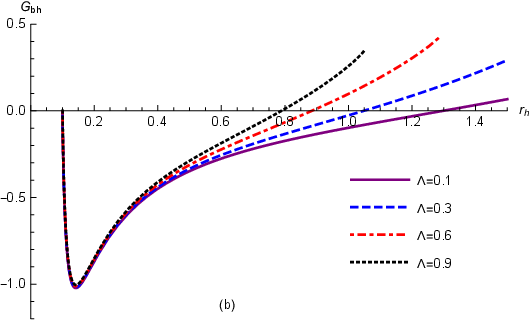}
\caption{Plot of (a) $G_{bh}$ versus $r_h$ with varying charge $Q$ (left plot) and $\lambda=0.4,a_k=0.1,c_t=1.5, c_r=1,\Lambda=0.2 $ and (b) $G_{bh}$ versus $r_h$ with varying parameter $\Lambda$ (right plot) and  $\lambda=0.4,Q=0.1,a_k=0.1,c_t=1.5, c_r=1$. }
\end{figure}

\begin{figure}[h!]
\centering
\includegraphics
{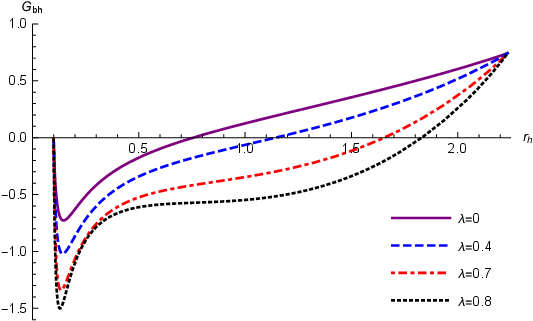}
\caption{Plot of $G_{bh}$ vs $r_h$ with varying parameter $\lambda$ and $Q=0.1,\Lambda=0.2, a_k=0.1, c_t=1.5, c_r=1$.}
\end{figure}

 \begin{table}[h!]

{\begin{tabular}{ |c| c| c| c| c|}
\hline
 $Q$ & Range of unstability & Range of stability & \multicolumn{2}{c|}{Modified Gibbs Free Energy}\\
 \cline{4-5}
 &  &  &   $r_h$ & minimum value \\ \hline
$0.10$ & $0.1001<r_h<1.1420$ & $1.1420<r_h<2.2338$ & $0.1452$ & $-1.0083$\\
\hline 
$0.15$ & $0.1503<r_h<1.1416$ & $1.1416<r_h<2.2338$ & $0.2210$ & $-0.6518$\\
\hline
$0.20$ & $0.2008<r_h<1.1410$ & $1.1410<r_h<2.2338$ & $0.2934$ & $-0.4792$\\
\hline
$0.25$ & $0.2517<r_h<1.1401$ & $1.1401<r_h<2.2338$ & $0.3645$ & $-0.3728$\\
\hline

\end{tabular}}
\caption{$G_{bh}$ vs $r_h$ with variable charge $Q$ and $\lambda=0.4,\Lambda=0.2, a_k=0.1, c_t=1.5, c_r=1$. }
\end{table}

 \begin{table}[h!]

{\begin{tabular}{ |c| c| c| c| c|}
\hline
 $\Lambda$ & Range of unstability & Range of stability & \multicolumn{2}{c|}{Modified Gibbs Free Energy}\\
 \cline{4-5}
 &  &  &   $r_h$ & minimum value \\ \hline
$0.1$ & $0.1001<r_h<1.2972$ & $1.2972<r_h<3.1607$ & $0.1479$ & $-1.01905$\\
\hline 
$0.3$ & $0.1001<r_h<1.0472$ & $1.0472<r_h<1.8230$ & $0.1478$ & $-1.01456$\\
\hline
$0.6$ & $0.1001<r_h<0.8823$ & $0.8823<r_h<1.2871$ & $0.1480$ & $-1.00778$\\
\hline
$0.9$ & $0.1001<r_h<0.7853$ & $0.7853<r_h<1.0493$ & $0.1481$ & $-1.00092$\\
\hline
\end{tabular}
}
\caption{$G_{bh}$ vs $r_h$ with variable $\Lambda$ and $Q=0.1,\lambda=0.4, a_k=0.1, c_t=1.5, c_r=1$. }
\end{table}
 \begin{table}[h!]

{\begin{tabular}{ |c| c| c| c| c|}
\hline
 $\lambda$ & Range of unstability & Range of stability & \multicolumn{2}{c|}{Modified Gibbs Free Energy}\\
 \cline{4-5}
 &  &  &   $r_h$ & minimum value \\ \hline
$0$ & $0.1001<r_h<0.7541$ & $0.7541<r_h<2.2338$ & $0.1484$ & $-0.72143$\\
\hline 
$0.4$ & $0.1001<r_h<1.1407$ & $1.1407<r_h<2.2338$ & $0.1452$ & $-1.00831$\\
\hline
$0.7$ & $0.1001<r_h<1.6467$ & $1.6467<r_h<2.2338$ & $0.1390$ & $-1.30971$\\
\hline
$0.8$ & $0.1001<r_h<1.8207$ & $1.8207<r_h<2.2338$ & $0.1345$ & $-1.44585$\\
\hline
\end{tabular}
}
\caption{$G_{bh}$ vs $r_h$ with variable $\lambda$ and $Q=0.1, \Lambda=0.2 ,a_k=0.1, c_t=1.5, c_r=1$.}
\end{table}


\subsection{Helmholtz free energy}
The Legendre transformation of the internal energy gives the Helmholtz free energy. It measures the amount of work which can be extracted from a system. The Helmholtz free energy function is constant if a system gets its reversible equilibrium state. In the presence of thermal fluctuation, the Helmholtz free energy of RNdS black hole can be calculated by the relation $F=G-P V$ [30]. Then the original and the modified Helmholtz free energy are
\begin{eqnarray}
F_{BH}&=&\frac{r_h}{4}\left(1-\frac{Q^2}{r_h^2}+\Lambda r_h^2\right)
\end{eqnarray}
and
\begin{eqnarray}
F_{bh}&=& \frac{r_h}{2}\left(1-\frac{Q^2}{r_h^2}\right)-\frac{1}{4L_1\pi r_h}\left(1-\frac{Q^2}{r_h^2}-\Lambda r_h^2\right)\Big[\pi r_h^2+\frac{1}{2}log (16 \pi L_1^2)\cr&&-log\left(1-\frac{Q^2}{r_h^2}-\Lambda r_h^2\right)\Big]
\end{eqnarray} 
respectively. It is noted that the original Helmholtz free energy of RNdS black hole is bigger than the modified Helmholtz free energy due to the presence of logarithmic term and Lorentz violation theory. The range of event horizon of original Helmholtz free energy will take all the values from zero to infinity but the range of event horizon of modified Helmholtz free energy will take all the values from $0.1$ to $2.23$ if $Q=0.1$ and $\Lambda=0.2$. This indicates that the range of event horizon of modified Helmholtz free energy also depends on charge of the black hole and cosmological constant $\Lambda$. 
The graphs of original and modified Helmholtz free energy are depicted in Figs. 9, 10 and 11 respectively. The behavior of the Helmholtz free energy is the same as that of Gibbs free energy. The graph of the original Helmholtz free energy is different from the graph of modified Helmholtz free energy due to quantum fluctuation and Lorentz violation theory. For different values of charge $Q$ and cosmological constant $\Lambda$, the positions of event horizon and its minimum values of the graph $10(a)$ and $10(b)$ are shown in Tables $(4)$ and $(5)$ respectively.  There is a equilibrium state of maximum entropy at the minimum condition of Helmholtz free energy. 
The graph of $F_{bh}$ is depicted with the radius of event horizon with the variation of Lorentz violation parameter $\lambda$. The modified Helmholtz free energy attains equilibrium position at $F_{bh}=1.11465$ for different values of parameter $\lambda$. The minimum values of $F_{bh}$ with different values of $\lambda$ are also shown in Table 6.  In the absence of quantum corrections and Lorentz violation theory, the figures of $10(a)$ and $10(b)$ coincide with figures $9(a)$ and $9(b)$ respectively. Figs. 11 is due to Lorentz violation theory only. This shows that the Helmholtz free energy is modified due to Lorentz violation theory.
\begin{figure}[h!]
\centering
\includegraphics[width=205pt,height=170pt]{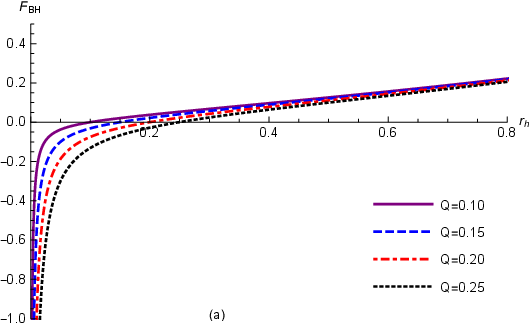}
\hfill
\includegraphics[width=205pt,height=170pt]{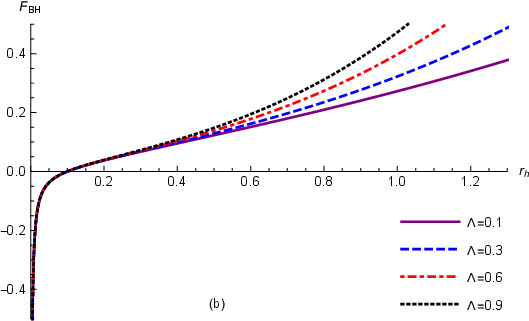}
\caption{Plot of $F_{BH}$ vs $r_h$ with varying charge $Q$ (left plot) and with varying $\Lambda$ (right plot) }
\end{figure}
 \begin{figure}[h!]
\centering
\includegraphics[width=205pt,height=170pt]{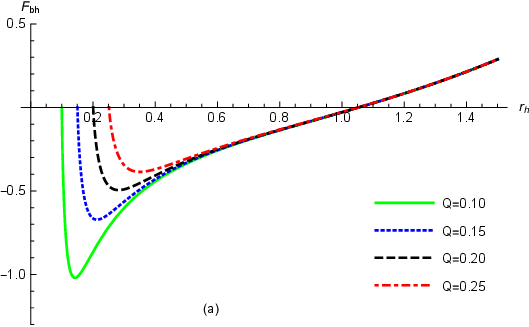}
\hfill
\includegraphics[width=205pt,height=170pt]{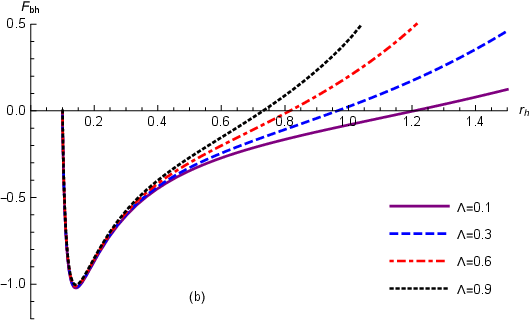}
\caption{Plot of $F_{bh}$ versus $r_h$ with varying charge, $Q$ (left plot) and cosmological constant, $\Lambda$ (right plot)respectively. Here, we take (a)  $\Lambda=0.2, a_k=0.1, c_t=1.5, c_r=1, \lambda=0.4$ and (b) $\lambda=0.4, Q=0.1, a_k=0.1, c_t=1.5, c_r=1$.}
\end{figure}
\begin{figure}[h]
\centering
\includegraphics{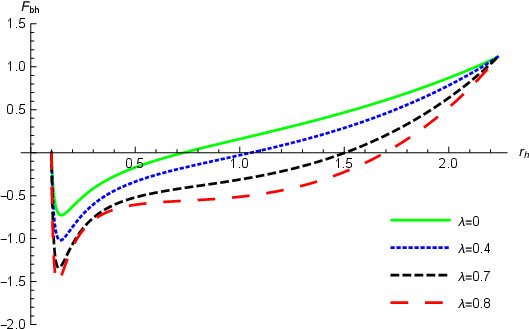}
\caption{Plot of $F_{bh}$ versus radius of event horizon, $r_h$ with varying parameter $\lambda$ and $Q=0.1,\Lambda=0.2, a_k=0.1, c_t=1.5, c_r=1$.}
\end{figure}

\begin{center}
\centering
\begin{table}[h!]

%
\begin{center}
\begin{tabular}{ |c|  c| c|}
\hline
   \qquad $Q$  \qquad  \qquad &   \multicolumn{2}{c|}{ \qquad Modified Helmholtz Free Energy  \qquad   \qquad}\\
 \cline{2-3}
 &  \qquad  \qquad   $r_h$  \qquad   \qquad \qquad &  \qquad minimum value  \qquad  \qquad \\ \hline
$0.10$  & $0.1452$ & $-1.0190$\\
\hline 
$0.15$ &  $0.2171$ & $-0.6712$\\
\hline
$0.20$ &  $0.2882$ & $-0.4942$\\
\hline
$0.25$ &  $0.3579$ & $-0.3851$\\
\hline
\end{tabular}
\end{center}

\caption{$F_{bh}$ vs $r_h$ with variable charge $Q$ and $\Lambda=0.2, \lambda=0.4, a=0.1, c_t=1.5, c_r=1$.}
\end{table}
\end{center}
\begin{table}[h!]

\begin{center}
\begin{tabular}{ |c|  c| c|}
\hline
   \qquad $\Lambda$  \qquad  \qquad &   \multicolumn{2}{c|}{ \qquad Modified Helmholtz Free Energy  \qquad   \qquad}\\
 \cline{2-3}
 &  \qquad  \qquad   $r_h$  \qquad   \qquad \qquad &  \qquad minimum value  \qquad  \qquad \\ \hline
$0.1$  & $0.1451$ & $-1.0213$\\
\hline 
$0.3$ &  $0.1452$ & $-1.0167$\\
\hline
$0.6$ &  $0.1454$ & $-1.0098$\\
\hline
$0.9$ &  $0.1455$ & $-1.0028$\\
\hline
\end{tabular}
\end{center}
\caption{$F_{bh}$ vs $r_h$ with variable $\Lambda$ and $Q=0.1,\lambda=0.4, a_k=0.1, c_t=1.5, c_r=1$.}
\end{table}
\begin{table}[h!]

\begin{center}
\begin{tabular}{ |c|  c| c|}
\hline
   \qquad $\lambda$  \qquad  \qquad &   \multicolumn{2}{c|}{ \qquad Modified Helmholtz Free Energy  \qquad   \qquad}\\
 \cline{2-3}
 &  \qquad  \qquad   $r_h$  \qquad   \qquad \qquad &  \qquad minimum value  \qquad  \qquad \\ \hline
$0$ &  $0.1504$ & $-0.7256$\\
\hline 
$0.4$ &  $0.1478$ & $-1.0167$\\
\hline
$0.7$ &  $0.1422$ & $-1.3299$\\
\hline
$0.8$  & $0.1378$ & $-1.4815$\\
\hline
\end{tabular}
\end{center}
\caption{$F_{bh}$ vs $r_h$ with variable $\lambda$ and $Q=0.1,\Lambda=0.2, a_k=0.1, c_t=1.5, c_r=1$.}
\end{table}
\section{Conclusion}
 In this paper, we investigate the tunneling of fermions near the event horizon of RNdS black hole using Dirac equation under Lorentz violation theory. The thermodynamic quantities such as entropy, Hawking temperature, heat capacity, Gibbs free energy and Helmholtz free energy are also studied by taking cosmological constant as thermodynamic variable and its conjugate quantity as thermodynamic volume under the Lorentz violation theory. From the Fig. 1, the Hawking temperature is modified for RNdS black hole due to Lorentz violation theory but the original and modified Hawking temperature intersect at the points $r_h=0.100303$ and $r_h=1.28709$. The graph of heat capacity of RNdS black hole is drawn in Fig. 2. There is a phase transition of RNdS black hole at $r_h=0.1716943$. It is noted that the position of phase transition and critical exponents are not affected by the Lorentz violation theory. These thermodynamic properties are related to the Lorentz violation parameter $\lambda$ but the equation of state derived from these thermodynamic quantities does not depend on Lorentz violation parameter $\lambda$ which is equivalent to the Einstein gravity when $\lambda=0$. We also observe that the critical pressure and volume are independent of Lorentz violation theory but the critical temperature depends on Lorentz violation parameter $\lambda$. \\ 
  A constant term $L_1$ appears in the expressions of  $\rho$ due to Lorentz violation theory. If $0<L_1<1$, it holds the inequality  $\rho<3/8$. When $1<L_1<\infty$, it obeys the inequality $\rho>3/8$. If $L_1=1$, $\rho=3/8$ which is universal constant for the van der Waals liquid-gas system. The first order phase transition from small black hole to large black hole which is equivalent to van der Waal liquid-gas system is derived. The entropy of black hole increases due to Lorentz violation theory but decreases due to quantum gravity correction. The path of the modified entropy of black hole due to Lorentz violation theory is similar to semi-like parabola whereas the path of original entropy is a parabola passing through the origin. Both the Gibbs free energy and Helmholtz free energy decrease due to the presence of Lorentz violation parameter $\lambda$.  The minimum values of Gibbs free energy and Helmholtz free energy of RNdS black hole are explored for different values of charge $Q$, cosmological constant, $\Lambda$ and Lorentz violation parameter, $\lambda$. From the graph of Gibbs free energy, it is noted that the numbers of Hawking-Page transition points increase due to Lorentz violation theory. For different values of charge $Q$ and cosmological constant $\Lambda$, the RNdS black hole is in radiative phase.
 \section*{Acknowledgements}
 The first author acknowledges the Department of Science and Technology (DST), New Delhi, India for giving financial support(Grant No. IF190759). 
\section*{Declaration of Competing Interest}
The authors declare that they do not have any conflict of interest.
 
\end{document}